\documentclass[%
 reprint,
 amsmath,amssymb,
 aps,
]{revtex4-2}
\usepackage[utf8]{inputenc}
\usepackage{graphicx}% for figures
\usepackage{dcolumn}% Align tables in the decimal number
\usepackage{bm}% bold math symbols
\usepackage{float}
\usepackage{footnote}
\usepackage{outlines}
\usepackage{amsmath}
\usepackage{subfigure}
\usepackage{hyperref}
\usepackage{comment}
\usepackage{amssymb}
\hypersetup{
    colorlinks=true,
    citecolor=blue,
    }

\begin{document}
%\title{}
\title{Digital quantum simulation of an extended Agassi model: Using   machine learning to disentangle its phase-diagram}

\author{Álvaro Sáiz$^1$, José-Enrique García-Ramos$^{2,3}$, José Miguel Arias$^{2,4}$, Lucas Lamata$^{2,4}$, Pedro Pérez-Fernández$^{1,2}$}
\affiliation{
    $^1$Departamento de Física Aplicada III\text{,} Escuela Técnica Superior de Ingeniería, Universidad de Sevilla, 41092 Sevilla, Spain.\\
    $^2$Instituto Carlos I de F\'{\i}sica Te\'orica y Computacional, Universidad de Granada\text{,} Fuentenueva s/n\text{,} 18071 Granada\text{,} Spain \\
    $^3$Departamento de  Ciencias Integradas y Centro de Estudios Avanzados en F\'isica\text{,} Matem\'atica y Computaci\'on, Universidad de Huelva,
21071 Huelva, Spain \\
    $^4$Departamento de F\'isica At\'omica, Molecular y Nuclear, Facultad de F\'isica\text{,} Universidad de Sevilla, Apartado 1065, E-41080 Sevilla, Spain
} 

\date{\today}

\begin{abstract}
  \begin{center}
    \rule{15cm}{1pt} \\
  \end{center}  
  A digital quantum simulation for the extended Agassi model is proposed using a quantum platform with eight trapped ions. The extended Agassi model is an analytically solvable model including both short range pairing and long range monopole-monopole interactions with applications in nuclear physics and in other many-body systems. In addition, it  owns a rich phase diagram with different phases and the corresponding phase transition surfaces. The aim of this work is twofold: on one hand, to propose a quantum simulation of the model at the present limits of the trapped ions facilities and, on the other hand, to show how to use a machine learning algorithm on top of the quantum simulation to accurately determine the phase of the system. Concerning the quantum simulation, this proposal is scalable with polynomial resources to larger Agassi systems. Digital quantum simulations of nuclear physics models assisted by machine learning may enable one to outperform the fastest classical computers in determining fundamental aspects of nuclear matter.
  \begin{center}
    \rule{15cm}{1pt} \\
  \end{center}
  
\end{abstract}

\maketitle
%\tableofcontents

\section{Introduction}
\label{sec-intro}
Realistic physical systems, in general, and nuclear physics problems, in particular, are usually too complex to be solved exactly. Thus, many-body approximations have to be used to get reasonable approximate results. Frequently, the complexity of the problem is such that even the use of standard many-body approximations faces insuperable difficulties. To tackle this issue, algebraic models, either for fermions or bosons, having only few degrees of freedom have been developed. These can be solved analytically in some particular scenarios and numerically in general cases, even for very large systems. These facts make them very useful to check approximations and to provide solid references to test more complex and fundamental calculations. Among this type of algebraic models, with relevance in Nuclear Physics, one can mention: the Elliot SU(3) model \cite{ElliotSU3}, which played a major role in the explanation of rotational structure emergence from single particle degrees of freedom in nuclei, the Interacting Boson Model (IBM) for the study of low-lying collective states in medium and heavy nuclei \cite{IBosonM}, or the Lipkin-Meshkov-Glick model (LMG) \cite{LMG} introduced first to check approximate many-body techniques in nuclear models with many nucleons interacting via a monopole-monopole force. Those are examples of simple models that have been able to permeate other fields of physics, e.g., condensed matter, quantum optics or molecular physics, due to its simplicity and ability to capture the key aspects of a large variety of physical systems. 

The Agassi model was proposed as a simple and exactly solvable version of the far more complex pairing-plus-quadrupole model of nuclear physics \cite{AGASSI196849}. It includes a combination of a two-level pairing model and a monopole-monopole interaction such as the one of the LMG model. The extended Agassi model was introduced in Ref. \cite{Garc_a_Ramos_2018} and its main difference with respect to the original model is the addition of a more general pairing interaction. The model presents a rich quantum phase diagram and, specifically, gives rise to quantum phase transitions of different nature. This makes the study of the extended Agassi model of much interest in the field of Quantum Phase Transitions (QPTs), in particular, to test new techniques able to identify the phase of the system through the study of appropriate observables.

Quantum simulations provide a fast-developing and powerful tool to provide realizations of countless quantum systems of great interest, such as spin models, quantum chemistry, as well as QPTs \cite{SUZUKIQPT}. In the near future, quantum simulators should be able to outperform classical computers and solve previously intractable problems. Consequently, many experimental setups are being proposed to validate the feasibility to deal with different physical models \cite{Tacchino_2019}. The usefulness of the Agassi model and its relevance in a variety of fields, including that of QPTs, motivates its quantum simulation. Furthermore, tools from quantum information have been recently used, as the quantum discord, to explore the phases in this model to gain insight about its structure \cite{Faba21}.

The quantum simulation of the original Agassi model has been already proposed for a system of trapped ions with four sites in \cite{perezfernandez2021quantum}, and the quantum phase diagram of the extended Agassi model has been extensively studied through a mean-field formalism \cite{GRamos2019Agassi}. In the present work, the quantum simulation of the extended Agassi model with eight sites in a system of trapped ions is worked out, as a natural followup to these previous studies. 
%%%%%NEW 
Note that the analysis is performed in a classical computer and the exact results are compared with the ones from a Trotter expansion which is used to simulate the quantum computer.  
% This poses a new challenge, as the mean field formalism is only valid for large-N, where N is the number of sites.
In the case of the standard Agassi model with four sites, its quantum simulation only presents two phases: a symmetric and a broken symmetry phase. However, the extended Agassi model with eight sites presents a richer phase diagram with five different quantum phases. This makes the model specially issued to be studied for two reasons: i) it is at the limit of what the current resources in facilities with trapped ions could do and ii) its rich phase diagram allows evaluating and validating the use of machine learning methods to identify the phase of the system.
%%%%%%%%%%%%%%%%%%%%%%%%%%%%%%%%%%%%%%%%%%%%%%%%%%%%%%%%
%that, because of finite-N effects, present a fuzzy limits. % bleed into each other, making the QPTs diffuse and requiring a more detailed study.

The phase/shape of a system is a property of its ground state. However, in Ref.~\cite{perezfernandez2021quantum} it was proved that it can also be determined through the time evolution study of selected operator matrix elements acting on a non eigenstate of the system. However, it is not always obvious how to determine the phase of the system through this procedure.
We can find an alternative solution to this challenge using another technique that has seen a fast growth in the last decade, namely, machine learning. Thanks to its great versatility, it has been tested with large success in many and diverse fields of science, such as medicine \cite{DeepLung}, biology \cite{AlphaFold} and astrophysics \cite{RArjonaML}. Machine learning algorithms are specially suitable to identify features that are not immediately apparent and they work as ideal classifiers \cite{oshea2015introduction}. In this paper, two different machine learning approaches are presented to aid in the determination of the system phase: one simple approach using a Multi-Layer Perceptron (MLP) and another more complex and precise approach, using a Convolutional Neural Network (CNN). Both are able to accurately predict the quantum phases of the system just taking as an input the quantum simulation of the time evolution of an appropriate operator.

The paper is organized as follows: in Section \ref{sectII}, the extended Agassi model Hamiltonian is presented and then it is written in terms of the Pauli matrices using the Jordan-Wigner spin mapping approach. In section \ref{sectIII}, the feasibility of a digital quantum simulation of the model is studied through its Trotterized dynamics that can be implemented efficiently with trapped-ion systems. In Section \ref{sectIV}, a brief study of the QPTs of the model is presented. In Section \ref{sectV}, the deep learning techniques, used for the determination of the quantum phase are presented and their results are reported. Finally, in Section \ref{sectVI}, the conclusions of the paper are summarized.

\section{Spin mapping of the extended Agassi model} \label{sectII}

The Agassi model is a two-level system in which the upper level is labeled as $\sigma = 1$ and the lower level as $\sigma = -1$. Each level has $\Omega$ degeneracy, being $\Omega$ an even number \cite{AGASSI196849}. One can then introduce $j$ such that $\Omega = 2j$.
Furthermore, in order to label the states within a given level, it is introduced a magnetic quantum number $m = \pm1, \pm2, \ldots, \pm \Omega /2$. In Ref.~\cite{Garc_a_Ramos_2018}, an extended Agassi model  was considered, which includes a more general pairing interaction with the additional term $-2hA_{0}^{\dagger}A_{0}$ (see Eq.~(\ref{H-eAm})). The Hamiltonian for the extended Agassi model is written as

%The Agassi model \cite{AGASSI196849} is a two-level system with each of the levels having a $\Omega$ degeneracy. The labeling for the single particle states are $\sigma = 1$ for the upper level and $\sigma= -1$ for the lower one, with the magnetic quantum number $m = \pm1, \pm2, \ldots, \pm \Omega /2$, which labels the states within a given level. Consequently, $\Omega$ is an even number and it is possible to introduce $j$ such that $\Omega = 2j$. In Ref.~\cite{Garc_a_Ramos_2018}, an extended Agassi model  was considered, which includes a more general pairing interaction with the additional term $-2hA_{0}^{\dagger}A_{0}$ (see Eq.~(\ref{H-eAm})). The Hamiltonian for the extended Agassi model is written as 
\begin{equation}
\begin{aligned}
H =\varepsilon J^{0}&-g\sum_{\sigma,\sigma^{\prime}=-1,1}^{}A_{\sigma}^{\dagger}A_{\sigma^{\prime}} \\ & -\frac{V}{2}\left[\left(J^{+}\right)^{2}+\left(J^{-}\right)^{2}\right]-2hA_{0}^{\dagger}A_{0}.
\end{aligned}
\label{H-eAm}
\end{equation}

The operators of the Hamiltonian are all defined in terms of fermion creation and annihilation operators,
\begin{align}
{J^{+}\,}&={\,\sum_{m=-j}^{j}c_{1,m}^{\dagger}c_{-1,m}\,=\,\left(J^{-}\right)^{\dagger}},\\
{J^{0}\,}&={\frac{1}{2}\,\sum_{m=-j}^{j}\left(c_{1,m}^{\dagger}c_{1,m}\,-\,c_{-1,m}^{\dagger}c_{-1,m}\right)},\\
{\,A_{1}^{\dagger}}&={\sum_{m=1}^{j}c_{1,m}^{\dagger}c_{1,-m}^{\dagger}} = (A_1)^\dagger,\\
{\,A_{-1}^{\dagger}}&={\sum_{m=1}^{j}c_{-1,m}^{\dagger}c_{-1,-m}^{\dagger}} = (A_{-1})^\dagger,\\
\,A_0^\dagger &= \sum_{m=1}^j \left(c_{-1,m}^{\dagger}c_{1,-m}^{\dagger} - c_{-1,-m}^{\dagger}c_{1,m}^{\dagger} \right) = (A_0)^\dagger,
\end{align}
where $c_{\pm1,m}^\dagger$ and $c_{\pm1,m}$ are the fermion creation and annihilation operators, respectively.

Using the Jordan-Wigner spin mapping approach \cite{Batista_2001,JordanWigner}, the Hamiltonian of the extended Agassi model can be mapped into Pauli matrices, which can later be experimentally implemented into a digital quantum simulator, as it will be discussed in the following section. In the following, a system with eight sites ($j=2$) will be considered. Hence, to simplify the notation, the fermions are re-labeled as:
% \begin{equation}
% \begin{aligned}
% {c_{1,2}\,\,\,\,\to\,c_{1}\,\,\,,\,\,\,\,\,\,c_{-1,2}\,\,\,\to\,c_{5}},\\
% {c_{1,1}\,\,\,\,\to c_{2}\,\,\,\,,\,\,\,\,\,\,c_{-1,1}\,\,\,\to\,c_{6}},\\
% {c_{1,-1}\to c_{3}\,\,\,\,,\,\,\,\,\,\,c_{-1,-1}\to c_{7}},\\
% {c_{1,-2}\to c_{4}\,\,\,\,,\,\,\,\,\,\,c_{-1,-2}\to c_{8}},	
% \end{aligned}    
% \end{equation}
\begin{equation}
\begin{aligned}
c_{1,2} &\to \,  c_{1}\,,\qquad c_{-1,2} &\to c_{5},\\
c_{1,1} &\to \,  c_{2}\,,\qquad c_{-1,1} &\to c_{6},\\
c_{1,-1} &\to \, c_{3}\,,\qquad c_{-1,-1} &\to c_{7},\\
c_{1,-2} &\to \, c_{4}\,,\qquad c_{-1,-2} &\to c_{8},	
\end{aligned}    
\end{equation}
and the corresponding re-labeling for the creation operators holds. The spin image of the above fermions is
\begin{equation}
\begin{aligned}
{c_{i}^{\dagger}\,}&={\,I_{1}\otimes\,...\,\otimes I_{i-1}\otimes \sigma_{i}^{+}\otimes \sigma_{i+1}^{z}\otimes\,...\,\otimes \sigma_{N}^{z}\,},\\
{c_{i}\,}&={\,I_{1}\otimes\,...\,\otimes I_{i-1}\otimes \sigma_{i}^{-}\otimes \sigma_{i+1}^{z}\otimes\,...\,\otimes \sigma_{N}^{z}\,},	
\end{aligned}
\end{equation}
where $\sigma^\pm$ are written in terms of the Pauli matrices in the standard way
% \begin{equation}
%     \sigma^{x} = 
%     \begin{pmatrix}
%     0 & 1\\
%     1 & 0
%     \end{pmatrix}
%     \,\,,\,\,\sigma^{y}=\,\begin{pmatrix}
%     {0}&{-i}\\
%     {i}&{0}\\
%     \end{pmatrix}\,\,,\,\,\sigma^{z}\,=\,\begin{pmatrix}
%     {1}&{0}\\
%     {0}&{-1}\\
%     \end{pmatrix},
% \end{equation}
\begin{equation}
    \sigma^+ = \frac{\sigma^x + i\sigma^y}{2}, \ \ 
    \sigma^- = \frac{\sigma^x - i\sigma^y}{2},
\end{equation}
$\sigma^{x,y,z}$ are, respectively,

\begin{equation}
\sigma^x=\left(\begin{array}{c c} 0 & 1\\1 & 0 \end{array}
\right), \sigma^y=\left(\begin{array}{c c} 0 & -i\\i & 0 \end{array}
\right), \sigma^z=\left(\begin{array}{c c} 1 & 0\\0 & -1 \end{array}
\right),
\end{equation}
$\otimes$ stands for the Kronecker product and $I$ for the identity operator.

The Hamiltonian can be written in terms of strings of tensor products of Pauli matrices. Indeed, there are terms which only depend on the Pauli matrix $\sigma^z$, and there are others that need to be written in terms of the Pauli matrices $\sigma^x$ and $\sigma^y$, so that they can be mapped into the corresponding quantum gates. Once $\sigma^+$ and $\sigma^-$ are decomposed in terms of $\sigma^x$ and $\sigma^y$, each of the tensor product strings will give rise to $8$ different terms. The total number of terms scales polynomially with the number of sites.

\section{Feasibility of the implementation}
\label{sectIII}
Given a certain time-independent Hamiltonian, the evolution operator is written as
\begin{equation}
  U(t) = e^{-itH},
\end{equation}
considering $\hbar = 1$.

For the case of a Hamiltonian written as the sum of $k$ different non commuting parts, the evolution operator can be approached using the so-called Trotter expansion as,
\begin{equation}
  \label{trotter_expansion}
  U(t,n)_T = \left( \prod_k e^{-itH_k/n} \right)^n \approx U(t).
\end{equation}

The study of the many Hamiltonian terms arising from the Jordan-Wigner mapping shows that many of them commute with each other. Hence, this can be exploited by adding these terms together in a total of $8$ non-commuting groups ($k=8$ in Eq. (\ref{trotter_expansion})), saving computational resources. Each of these groups can then be implemented with trapped-ion systems using a combination of single and multiqubit (M{\o}lmer–S{\o}rensen) gates and simulation of the terms within each single group can be carried out sequentially without digital error \cite{PhysRevLett.113.220501,PhysRevLett.117.060504,CasanovaPRL2012}.

The accuracy of this approximation can be measured by its fidelity, defined as,
\begin{equation}
  F(t,n_T) = |\langle\phi| U(t,n_T)_T^\dagger U(t)|\phi\rangle|^2.
\end{equation}

This observable depends on time, $t$, and on the number of Trotter steps $n_T$. The considered initial state to which we apply the evolution will be along the rest of the paper $|\downarrow_1 \ \downarrow_2 \ \downarrow_3 \ \downarrow_4 \ \uparrow_5 \ \uparrow_6 \ \uparrow_7 \ \uparrow_8 \ \rangle$, where  $|\downarrow_i\rangle$ and $|\uparrow_i\rangle$ stand for the eigenvectors of $\sigma^z_i$ with eigenvalues $-1$ and $1$, respectively. Hence, the trial state is not a Hamiltonian eigenstate, except for $V = g = h = 0$.

\begin{figure}
  \centering
  \includegraphics[scale=0.35]{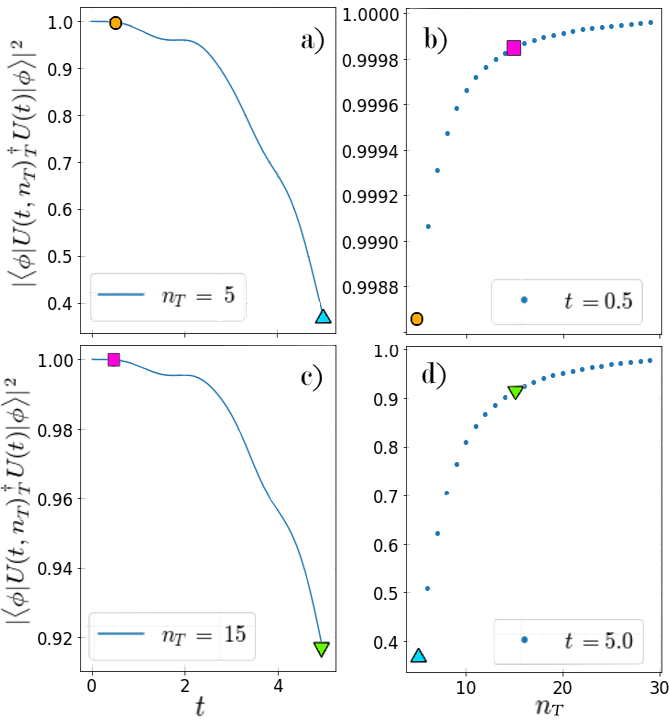}
  \caption{Fidelity as a function of $t$ for a) $n_T = 5$ and c) $n_T = 15$ and as a function of $n_T$ for b) $t=0.5$ and d) $t=5$. The initial state in all cases is $|\downarrow_1 \ \downarrow_2 \ \downarrow_3 \ \downarrow_4 \ \uparrow_5 \ \uparrow_6 \ \uparrow_7 \ \uparrow_8 \ \rangle$ and the Hamiltonian parameters are $\varepsilon = 1$, $g = 0.25$, $V = 0.25$ and $h = 0.25$. The highlighted points correspond to $n_T = 5, 15$ and $t = 0.5, 5$, respectively.}
  \label{fig:fidelity}
\end{figure}

Fig.~\ref{fig:fidelity} shows the fidelity as a function of both $t$ and $n_T$. In a previously studied case with $4$ sites ($j=1$), it was seen that already $n_T=5$ Trotter steps provided enough accuracy \cite{perezfernandez2021quantum}. Here, the fidelity drops faster as time increases and for $n_T = 5$ the approximation only remains accurate at small times ($t < 1$) (panel a)), but this drop is greatly reduced for $n_T = 15$ (panel c)). This is expected, since $8$ sites are considered and, therefore, there exist many more terms that need to be considered for the Trotterized Hamiltonian, which in turn means that a significantly higher amount of Trotter steps are required to attain the same accuracy in a given time range. In panels  b) and d) the value of the fidelity is depicted as a function of the Trotter steps, $n_T$, for two different values of time %The models presented in this paper will be working with times in the range of $t \approx 5$, which will require a relatively high amount of Trotter steps.

\begin{figure}
  \centering
  \includegraphics[scale=0.35]{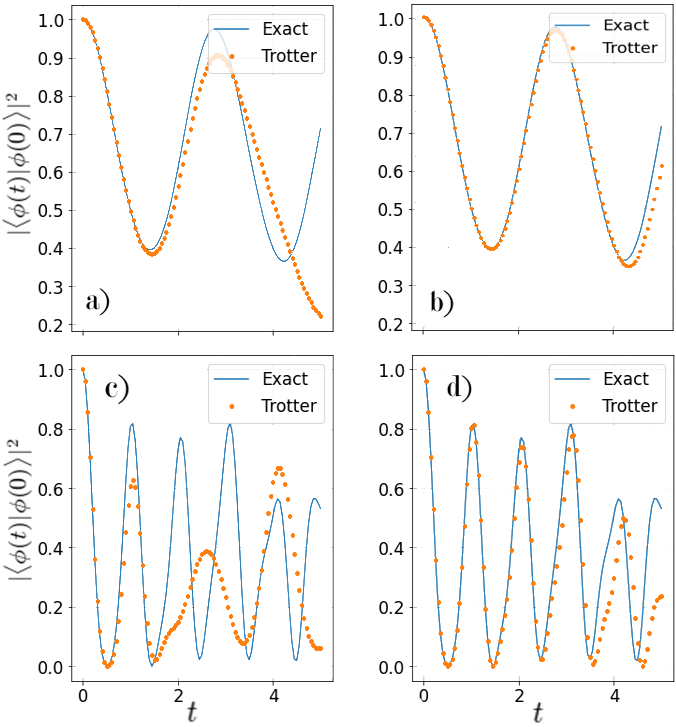}
  \caption{Survival probability $|\langle \phi (t) | \phi (0) \rangle | ^2$ as a function of $t$ using the exact and the Trotterized evolution operator. In all cases, the initial state is $|\downarrow_1 \ \downarrow_2 \ \downarrow_3 \ \downarrow_4 \ \uparrow_5 \ \uparrow_6 \ \uparrow_7 \ \uparrow_8 \ \rangle$ and $\varepsilon = 1$, $g = 0.5$ and $h = 1.5$. \textit{Top row:} $V=0.25$ and a) $n_T=5$ and b) $n_T=15$. \textit{Bottom row:} $V=1.5$ and c) $n_T=5$ and d) $n_T=30$.}
  \label{fig:survival}
\end{figure}

%The fidelity mainly drops with time, but it also depends on the value of all three parameters of the Hamiltonian: $g$, $V$ and $h$.
Fig.~\ref{fig:survival} shows as an example the survival probability, $|\langle \phi (t)| \phi (0)\rangle|^2$, computed for both the exact and the Trotterized evolution operator with different values of $V =0.25$  (top row) and $V= 1.5$ (bottom row) and $n_T = 5, \ 15$ (panels a) and b)) and $n_T = 5, \ 30$ (panels c) and d)), keeping constant $g = 0.5$ and $h = 1.5$. For a small value of $V=0.25$, a good fidelity can be achieved with only $n_T=5$ Trotter steps. However, this is not true in general. There are situations where an accurate study require a larger number of Trotter steps as it is seen in Fig.~\ref{fig:survival}, panels c) and d).

The implementation of the simulation is efficient, that is, it scales with polynomial resources. However, considering the need of two M{\o}lmer–S{\o}rensen gates for each tensorial product and two single-qubit gates for each $\sigma^x$ and $\sigma^y$ to be changed into a $\sigma^z$ and back (assuming $\sigma^z$ as the basis), around two thousand quantum gates would be needed to implement this model for $j=2$. It is important to note that the aggregated error from all the gates will then increase with the number of Trotter steps needed for a precise approximation. Nevertheless, the estimations seem to agree that this proposal should be achievable in the near future as a good progress is being made in scaling trapped ion systems to the hundreds or thousands of qubits \cite{ScalingIon,ScalingIon2}. Some approaches with promising results are the implementation of high fidelity fast gates \cite{2017fastgates,2020FastGates} and two dimensional ion traps, since the fidelities  in each dimension are independent from one another \cite{twodimIonTraps}. The use of a higher-order Trotter expansion would also help reducing the required number of Trotter steps for an accurate approximation \cite{Batista_2001}.

\section{Exploring phase transitions}
\label{sectIV}

The original Agassi model presents a rich quantum phase diagram with a symmetric, a parity broken phase and a superconducting one \cite{AgassiPhase1}. By adding the additional pairing interaction term of the extended Agassi model, the quantum phase diagram becomes even richer. In order to explore the phase diagram and the appearance of QPTs it is customary to rescale the parameters of the Hamiltonian (\ref{H-eAm}) by introducing the new parameters $\chi$, $\Sigma$ and $\Lambda$,
\begin{equation}
  V = \frac{\varepsilon \chi}{2j - 1}, \ \ \
  g = \frac{\varepsilon \Sigma}{2j - 1}, \ \ \
  h = \frac{\varepsilon \Lambda}{2j - 1}.
\end{equation}

\begin{figure}
    \centering \includegraphics[scale=0.3]{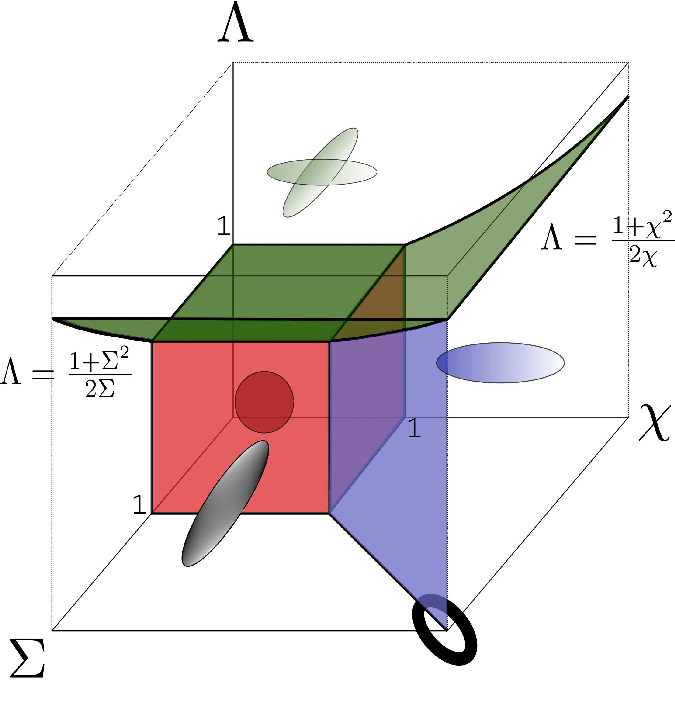}
    \caption{Graphical representation of the phase diagram of the extended Agassi Hamiltonian. Red vertical planes represent second order QPT surfaces. The green surface ($\Lambda = 1$ for $\chi < 1$ and $\Sigma < 1$, $\Lambda = \frac{1+\chi^2}{2 \chi}$ for $\chi > \Sigma$ and $\Lambda = \frac{1+\Sigma^2}{2 \Sigma}$ for $\chi < \Sigma$) and the blue vertical one ($\chi = \Sigma$ and $\Lambda < \frac{1 + \Sigma^2}{2 \Sigma}$) correspond to first order critical surfaces. Red sphere, blue oval, black oval, black thick oval and crossed green ovals correspond to the symmetric solution, the HF deformed solution, the BCS deformed solution, the closed valley solution and the HF-BCS deformed solution, respectively. Figure adapted from \cite{Garc_a_Ramos_2018}}.
    \label{fig:PhaseDiagram}
\end{figure}

The phase diagram of the extended Agassi model presents five different quantum phases, as shown in Fig.~\ref{fig:PhaseDiagram}. They appear in the following areas of the phase diagram (see Ref.~\cite{GRamos2019Agassi} for a detailed study of the quantum phases of the system):
\begin{itemize}
\item Symmetric or spherical solution: $\\ \chi < 1, \ \ \Sigma < 1, \ \ \Lambda < 1$.
\item Hartree-Fock (HF) deformed solution: $\\ \chi > 1, \ \ \Sigma < \chi, \ \ \Lambda < \frac{1 + \chi^2}{2\chi}$.
\item Bardeen–Cooper–Schrieffer (BCS) deformed solution: $ \chi < \Sigma, \ \ \Sigma > 1, \ \ \Lambda < \frac{1 + \Sigma^2}{2\Sigma}$.
\item Combined HF-BCS deformed solution: \\ $\Lambda > 1 \ \text{for} \ \chi < 1 \ \text{and} \ \ \Sigma < 1, \\ \ \ \Lambda > \frac{1 + \chi^2}{2\chi} \ \text{for} \ \chi > \Sigma > 1, \\ \ \ \Lambda > \frac{1 + \Sigma^2}{2\Sigma} \ \text{for} \ \Sigma > \chi > 1$.
\item Closed valley solution: $\chi = \Sigma, \ \ \Lambda < \frac{1 + \Sigma^2}{2\Sigma}$.
\end{itemize}

It is important to note that this phase diagram has been obtained using a mean-field approach, which is only valid for the large $N$ limit, where $N$ is the number of sites, and it allows to calculate the value of the order parameters as a function of the control parameters. Here, $N=8$ and therefore, finite-$N$ effects become relevant, causing the QPTs to be diffuse and harder to be distinguished. 

In previous studies, time evolution of a correlation function has been successfully used as a tool to explore the QPTs of the standard Agassi model with $j=1$ \cite{perezfernandez2021quantum}. In the rest of this work, we will use precisely this observable to study the phase diagram of the extended Agassi model with $j=2$. A correlation function between two given sites can be defined as,
\begin{equation}
    C_{\alpha,\beta}(i,j) = 
    \langle \sigma_i^{\alpha} \otimes \, \sigma_j^{\beta} \rangle -
    \langle \sigma_i^{\alpha} \rangle \, \langle \sigma_j^{\beta} \rangle,
\end{equation}
where $\alpha, \ \beta =  x, \ y \text{ or } z$. In our case, we will focus on a correlation function for sites $1$ and $2$ and $\alpha=\beta=z$
\begin{equation}
  \label{sigma12}
  C_z(1,2) \equiv \langle \sigma_1^{z} \otimes \, \sigma_2^{z} \rangle - \langle \sigma_1^{z} \rangle \, \langle \sigma_2^{z} \rangle.
\end{equation}  

\begin{figure*}
\begin{subfigure}
    \centering
    \includegraphics[scale=0.4]{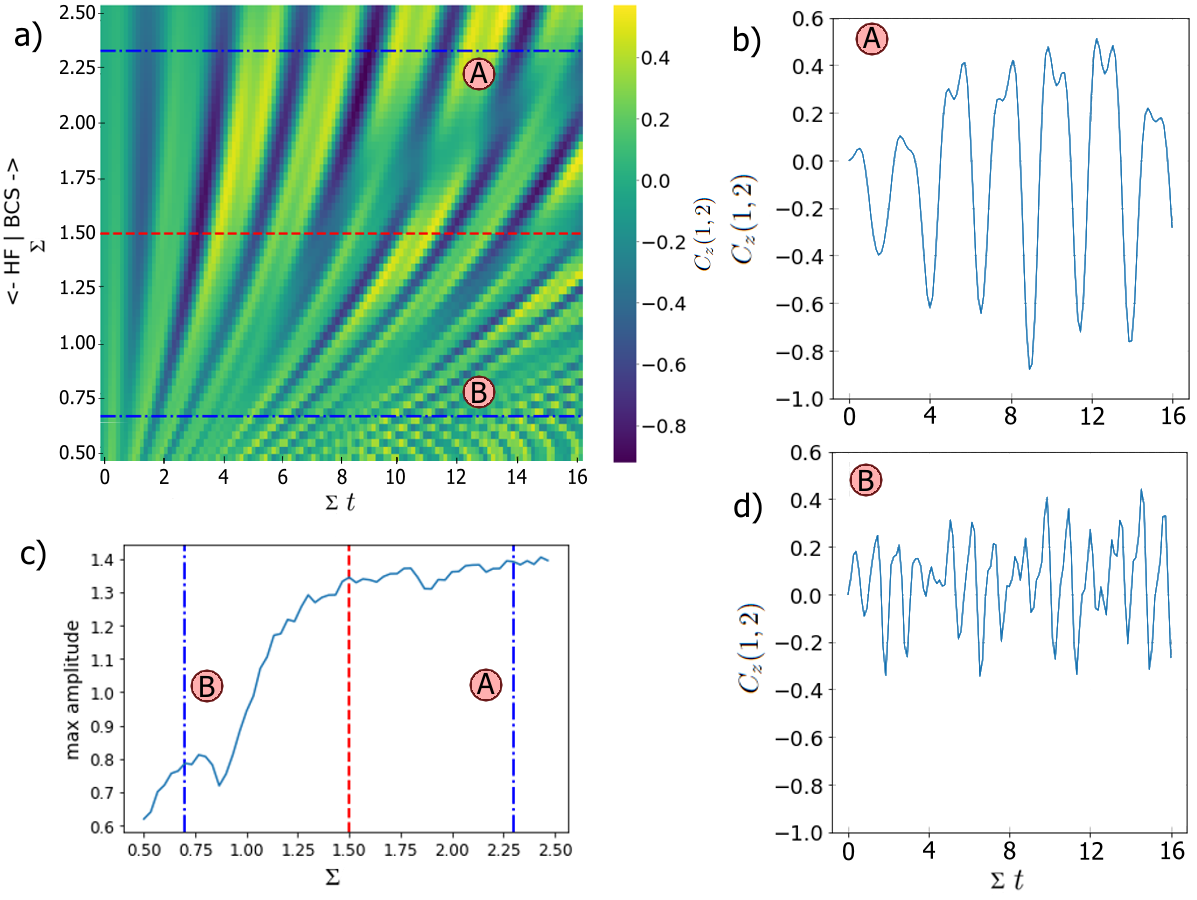}
    \caption{
    \textbf{a}) Time evolution of the correlation function $C_z(1,2)$ from $\Sigma = 0.5$ to $\Sigma=2.5$, with $\chi = 1.5$, $\Lambda=0$, and an initial state $|\downarrow_1 \ \downarrow_2 \ \downarrow_3 \ \downarrow_4 \ \uparrow_5 \ \uparrow_6 \ \uparrow_7 \ \uparrow_8 \ \rangle$. The phase transition point $\Sigma = \chi = 1.5$ is marked by a dashed red line and the points $\chi=1.5$, $\Sigma = 0.7$ (slice B) and $\chi=1.5$, $\Sigma = 2.3$ (slice A) with blue dashed lines. \textbf{c)} Maximum value of the amplitude of the oscillations of $C_z(1,2)$ as a function of $\Sigma$ value. \textbf{b}) Time evolution of $C_z(1,2)$ for $\Sigma = 2.3$ (slice A) and \textbf{d)} for $\Sigma = 0.7$ (slice B).
    }
    \label{fig:C12corr}
\end{subfigure}
\end{figure*}

Fig.~\ref{fig:C12corr} a) shows the systematics of the correlation function $C_z(1,2)$  as we move through a line of fixed $\chi = 1.5$ and $\Lambda = 0$ from the HF phase ($\Sigma < \chi$) to the BCS phase ($\Sigma>\chi$) as a function of time. The correlation function oscillates with time and, despite the fact that the transition is not apparent, if we move away from the critical surface, e.g., at slices A and B, we can see a clear difference in the amplitude of these oscillations. Hence, it increases as one moves towards the BCS phase, as shown in panel c) of Fig.~\ref{fig:C12corr}. In panels b) and d), the time evolution of the correlation function for the points A and B are depicted. These two panels clearly show how the amplitude and the period of the oscillation greatly change depending on the value of the control parameter. In Appendix \ref{sec-app-A}, other examples are gathered showing the variety of behaviors that one can get depending on the position in the phase diagram. 
%Similar behaviours can be appreciated for transitions between different phases, where other characteristics of the oscillations vary from one phase to another, namely, the amplitude, frequency and/or the mean value, as well as modulations of the amplitude that may appear. Some examples have been gathered in appendix \hyperlink{appendixB}{B}. 
Other correlation functions between different sites, applied to the same and/or different initial states, yield similar results (not shown in this work), except when they vanish, which proves that correlation functions are good observables to obtain information about the phase of the system without the need of computing the ground state.

Even in the case where differences between quantum phases can be appreciated in the time evolution, it is hard, and sometimes impossible, to visually identify the quantum phase of the system, specially when looking at a single point in the phase space. 
%For this reason, a machine learning procedure has been designed to identify the phase of the system. 
Hence, two machine learning models will be trained for quantum phase classification using the time dynamics of the correlation function $C_z(1,2)$. The benefit of using machine learning methods for classification is that, given the data to classify, the method can determine itself the optimal procedure and, moreover, it will be able to accurately label not previously considered cases.
%parameters to distinguish the set of input data (in this case $C_z(1,2)$) by the provided classification labels (in this case, the different quantum phases), even if those parameters have not a direct physical meaning.

\section{Machine learning methods for classification}
\label{sectV}
\subsection{Multi-Layer Perceptron}
\label{sectV-MLP}

A Multi-Layer Perceptron (MLP) is a supervised machine learning algorithm that learns a non-linear function which, given a set of features as input $X = x_1, x_2, \ldots, x_m$, is able to approximately predict a target output $y$. The algorithm learns this function via gradient descent, modifying its weights by minimizing a loss function based on how accurate the predictions $y\prime$ are in relation to the real target values $y$, which are known in the training process, as with any supervised method \cite{MLP}. Once optimized, it can reliably predict the target $y\prime \approx y$ for inputs for which the real target is not already known. 

%A Multi-Layer Perceptron (MLP) is a supervised machine learning algorithm that learns a function $F(\cdot): R^m \rightarrow R^o$ by training on a dataset, where $m$ is the dimension of the input and $o$ is the dimension of the output. Given a set of features $X = x_1, x_2, \ldots, x_m$ and a target $y$, it can learn a non-linear function {\it approximator} for either classification or regression by error propagation \cite{MLP}.

The implementation of this model is very simple thanks to the existing standardized machine learning software packages, e.g., \textit{scikit-learn} \cite{scikit-learn}, but first it is needed to define a set of training and test data for the model to learn with. For this purpose, we compute the exact as well as the Trotterized (using a number of Trotter steps $n_T=6$) time evolution of the correlation operator $C_z(1,2)$ for a large set of points throughout the phase diagram. We create a mesh of points with the parameters $\chi$, $\Sigma$ and $\Lambda$ in the interval $[0,2]$ with 21 steps each for a total of $9261$ points. For each of them, we compute the time dynamics for $100$ time steps. This information together with the knowledge of the phase diagram (see Section \Ref{sectIV}) feeds the model as part of the training process.

Once the dataset is ready, it is randomly split into two sets. One set, the training set, will be used to train the model, while the other, the test set, will be used to test the accuracy of the model. It is compulsory for the model to be tested with data not used during the training process to be sure that the algorithm is not simply remembering the already used data. Therefore, it is important to have enough data points for testing and to reach an acceptable metric, while keeping as many points as possible for training. In most cases, to use from $10$ to $20\%$ of data for testing is enough. In our case, we will use $10\%$ of data points for testing.

\textit{Scikit-learn} provides us with several optimization algorithms to perform training on the data. In this particular case, the default \textit{Adam} solver gives optimal results. \textit{Adam} solver is an optimization algorithm that is based on the stochastic gradient descent \cite{kingma2017adam}. It combines the advantages of two other extensions of the stochastic gradient descent known as Adaptive Gradient Algorithm (AdaGrad) and Root Mean Square Propagation (RMSProp). For a general overview on the theoretical and practical aspects of the learning algorithms based on stochastic gradient descent, see, e.g., Ref.~\cite{GradDescBottou2004}.

The model actually computes the probability for each quantum phase and classifies the data based on which one is the most probable. The classification with the MLP  correctly predicts the quantum phase in the $96.7\%$ of the points in the test data when trained and tested with the exact time evolution. When the MLP is trained and tested with the Trotterized time evolution, it has an accuracy of $93.4\%$ ($n_T=6$). These results are very valuable, since it proves that the quantum phase can be determined using the correlation function without the need of computing the ground state, but it also shows that they can be identified despite using the Trotter approximation with a relative small number of steps, for which the agreement with the exact results is limited. Note that reducing the number of Trotter steps, the aggregated error of the quantum gates is reduced too. 

\begin{figure*}
    \centering
    \includegraphics[scale=0.7]{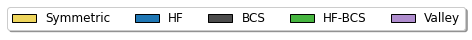}
    \begin{tabular}{c c}
        \vspace{3pt}
        \textbf{\ \ \ a)} $\Sigma = 0.5$  ;  $\Lambda = 0.5$ & \textbf{\ \ \ b)} $\chi = 0.5$  ;  $\Lambda = 0.5$ \\
        \hspace{-6pt}\begin{tabular}{c c}
            \vspace{-5pt}
            \text{\ \ \ \ \  Exact} & \text{Trotter} \\
            \includegraphics[scale=0.5,trim={8pt 0 0 0},clip]{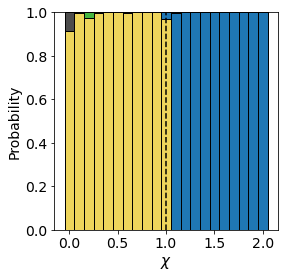} &
            \hspace{-5pt}\includegraphics[scale=0.5,trim={50pt 0 0 0},clip]{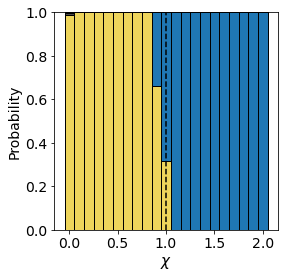}
        \end{tabular} &
        \hspace{-6pt}\begin{tabular}{c c}
            \vspace{-5pt}
            \text{\ \ \ \ \  Exact} & \text{Trotter} \\
            \includegraphics[scale=0.5,trim={8pt 0 0 0},clip]{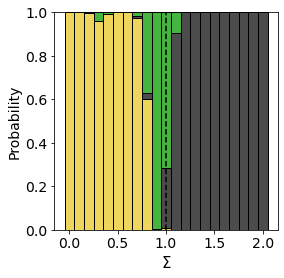} &
            \hspace{-5pt}\includegraphics[scale=0.5,trim={50pt 0 0 0},clip]{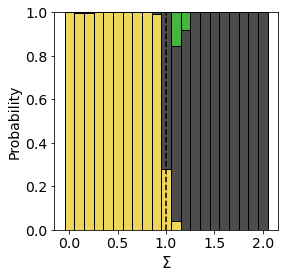}
        \end{tabular} \\
        \vspace{3pt}
        \textbf{\ \ \ c)} $\chi = 0.5$  ;  $\Sigma = 0.5$ & \textbf{\ \ \ d)} $\chi = 1.5$  ;  $\Lambda = 0.5$ \\
        \hspace{-6pt}\begin{tabular}{c c}
            \vspace{-5pt}
            \text{\ \ \ \ \  Exact} & \text{Trotter} \\
            \includegraphics[scale=0.5,trim={8pt 0 0 0},clip]{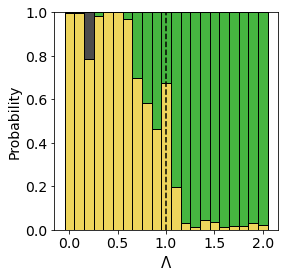} &
            \hspace{-5pt}\includegraphics[scale=0.5,trim={50pt 0 0 0},clip]{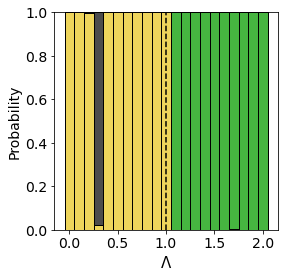}
        \end{tabular} &
        \hspace{-6pt}\begin{tabular}{c c}
            \vspace{-5pt}
            \text{\ \ \ \ \  Exact} & \text{Trotter} \\
            \includegraphics[scale=0.5,trim={8pt 0 0 0},clip]{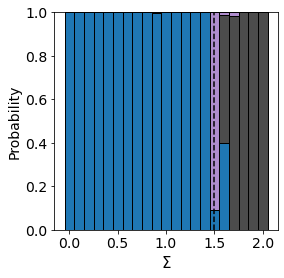} &
            \hspace{-5pt}\includegraphics[scale=0.5,trim={50pt 0 0 0},clip]{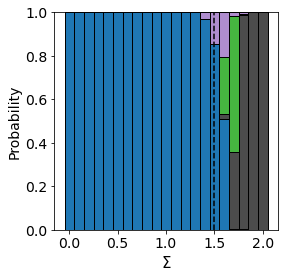}
        \end{tabular} \\
        \vspace{3pt}
        \textbf{\ \ \ e)} $\chi = 1.5$  ;  $\Sigma = 0.5$ & \textbf{\ \ \ f)} $\chi = 0.5$  ;  $\Sigma = 1.5$ \\
        \hspace{-6pt}\begin{tabular}{c c}
            \vspace{-5pt}
            \text{\ \ \ \ \  Exact} & \text{Trotter} \\
            \includegraphics[scale=0.5,trim={8pt 0 0 0},clip]{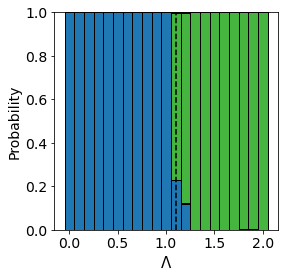} &
            \hspace{-5pt}\includegraphics[scale=0.5,trim={50pt 0 0 0},clip]{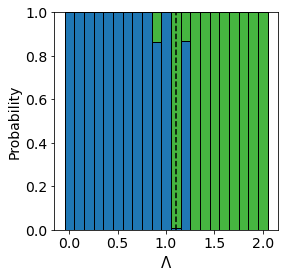}
        \end{tabular} &
        \hspace{-6pt}\begin{tabular}{c c}
            \vspace{-5pt}
            \text{\ \ \ \ \  Exact} & \text{Trotter} \\
            \includegraphics[scale=0.5,trim={8pt 0 0 0},clip]{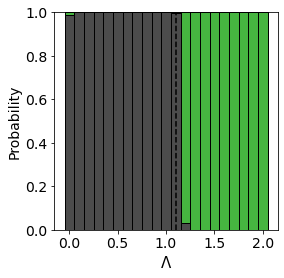} &
            \hspace{-5pt}\includegraphics[scale=0.5,trim={50pt 0 0 0},clip]{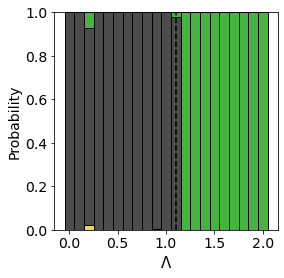}
        \end{tabular} \\
    \end{tabular}
    \caption{Quantum phase prediction of the system via the MLP method applied to the $C_z(1,2)$ correlation function. The different panels show the probability for the system of being in a given phase for specific values of the Hamiltonian parameters. Left panels correspond to the results using the exact time evolution, while right ones using the Trotter approach with $n_T=6$. In all cases, the initial state $|\downarrow_1 \ \downarrow_2 \ \downarrow_3 \ \downarrow_4 \ \uparrow_5 \ \uparrow_6 \ \uparrow_7 \ \uparrow_8 \ \rangle$ is used. Panel a) $\Sigma = 0.5$ and $\Lambda = 0.5$ moving from $\chi = 0$ (symmetric phase) to $\chi = 2$ (HF phase); panel b) $\chi = 0.5$ and $\Lambda = 0.5$ moving from $\Sigma = 0$ (symmetric phase) to $\Sigma = 2$ (BCS phase); panel c) $\chi = 0.5$ and $\Sigma = 0.5$ moving from $\Lambda = 0$ (symmetric phase) to $\Lambda = 2$ (combined HF-BCS phase); panel d) $\chi = 1.5$ and $\Lambda = 0.5$ moving from $\Sigma = 0$ (HF phase) to $\Sigma = 2$ (BCS); panel e) $\chi = 1.5$ and $\Sigma = 0.5$ moving from $\Lambda = 0$ (HF phase) to $\Lambda = 2$ (combined HF-BCS phase); panel f) $\chi = 0.5$ and $\Sigma = 1.5$ moving from $\Lambda = 0$ (BCS phase) to $\Lambda = 2$ (combined HF-BCS phase). The dashed black line in each panel marks the theoretical critical point.}
    \label{fig:MLPpred}
\end{figure*}

In Fig.~\ref{fig:MLPpred}, the results for the MLP method are presented through different trajectories in the parameter space of the model. Generally, the incorrect predictions occur near the critical points, which is expected, since the QPTs are not clearly defined because of finite-$N$ effects ($N=8$ in our case). %Despite some individual incorrect predictions, in most cases one can identify a transition from one predicted phase to another one around the point where the QPT actually occurs.
%It must be noted that to train the MLP one must give each point a single label. Thus, although in the
Note that at the critical points (denoted by a dashed line), two or three quantum phases can coexist, therefore, it is considered that the model correctly classifies the quantum phase of the system when it predicts either of the two phases, except in the case of the closed valley solution, where that phase is considered the only valid answer. This is done to ensure that the model learns to identify this specific phase instead of choosing any of the other two phases with which it coexists. 

Perhaps the most surprising case is, in fact, the one regarding this solution. In the results portrayed in Fig.~\ref{fig:MLPpred} d), the model is able to predict the quantum phase of the system as it crosses the closed valley solution at $\chi = \Sigma = 1.5$ when trained with the exact solution. As mentioned, the model is not completely accurate in some cases [as can be seen from Fig.~\ref{fig:MLPpred}c)-Trotter], but the QPTs are still fairly clear. 
%Even then, the results are remarkable for such a fast and easy to implement model. Moving forward, a more complex and more accurate machine learning model is presented.
In next section we will present a  more accurate machine learning model.

\begin{figure*}
    \centering
    \includegraphics[scale=0.7]{Figures/Convolution/legend.png}
    \begin{tabular}{c c}
        \vspace{3pt}
        \textbf{\ \ \ a)} $\Sigma = 0.5$  ;  $\Lambda = 0.5$ & \textbf{\ \ \ b)} $\chi = 0.5$  ;  $\Lambda = 0.5$ \\
        \hspace{-6pt}\begin{tabular}{c c}
            \vspace{-5pt}
            \text{\ \ \ \ \  Exact} & \text{Trotter} \\
            \includegraphics[scale=0.5,trim={8pt 0 0 0},clip]{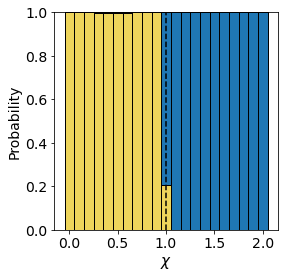} &
            \hspace{-5pt}\includegraphics[scale=0.5,trim={50pt 0 0 0},clip]{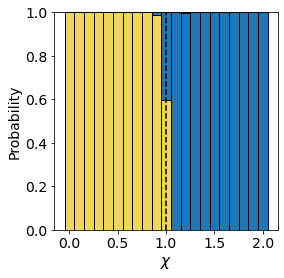}
        \end{tabular} &
        \hspace{-6pt}\begin{tabular}{c c}
            \vspace{-5pt}
            \text{\ \ \ \ \  Exact} & \text{Trotter} \\
            \includegraphics[scale=0.5,trim={8pt 0 0 0},clip]{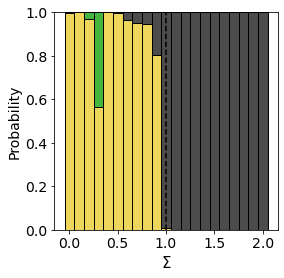} &
            \hspace{-5pt}\includegraphics[scale=0.5,trim={50pt 0 0 0},clip]{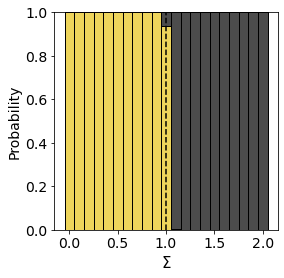}
        \end{tabular} \\
        \vspace{3pt}
        \textbf{\ \ \ c)} $\chi = 0.5$  ;  $\Sigma = 0.5$ & \textbf{\ \ \ d)} $\chi = 1.5$  ;  $\Lambda = 0.5$ \\
        \hspace{-6pt}\begin{tabular}{c c}
            \vspace{-5pt}
            \text{\ \ \ \ \  Exact} & \text{Trotter} \\
            \includegraphics[scale=0.5,trim={8pt 0 0 0},clip]{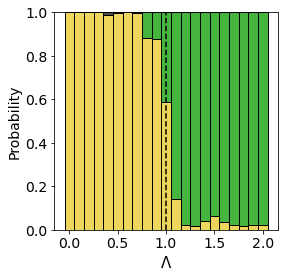} &
            \hspace{-5pt}\includegraphics[scale=0.5,trim={50pt 0 0 0},clip]{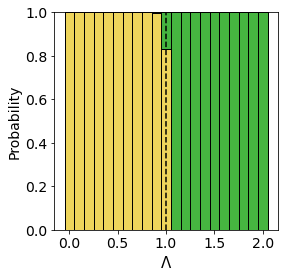}
        \end{tabular} &
        \hspace{-6pt}\begin{tabular}{c c}
            \vspace{-5pt}
            \text{\ \ \ \ \  Exact} & \text{Trotter} \\
            \includegraphics[scale=0.5,trim={8pt 0 0 0},clip]{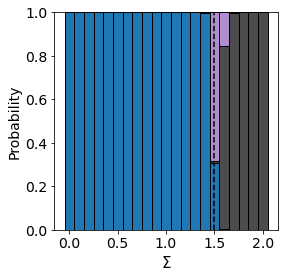} &
            \hspace{-5pt}\includegraphics[scale=0.5,trim={50pt 0 0 0},clip]{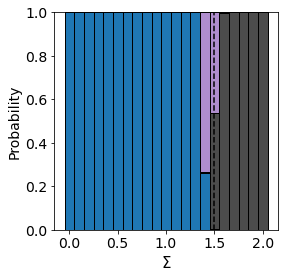}
        \end{tabular} \\
        \vspace{3pt}
        \textbf{\ \ \ e)} $\chi = 1.5$  ;  $\Sigma = 0.5$ & \textbf{\ \ \ f)} $\chi = 0.5$  ;  $\Sigma = 1.5$ \\
        \hspace{-6pt}\begin{tabular}{c c}
            \vspace{-5pt}
            \text{\ \ \ \ \  Exact} & \text{Trotter} \\
            \includegraphics[scale=0.5,trim={8pt 0 0 0},clip]{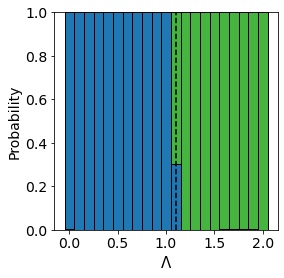} &
            \hspace{-5pt}\includegraphics[scale=0.5,trim={50pt 0 0 0},clip]{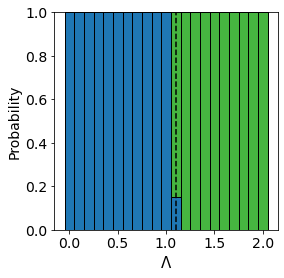}
        \end{tabular} &
        \hspace{-6pt}\begin{tabular}{c c}
            \vspace{-5pt}
            \text{\ \ \ \ \  Exact} & \text{Trotter} \\
            \includegraphics[scale=0.5,trim={8pt 0 0 0},clip]{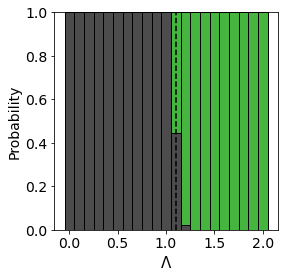} &
            \hspace{-5pt}\includegraphics[scale=0.5,trim={50pt 0 0 0},clip]{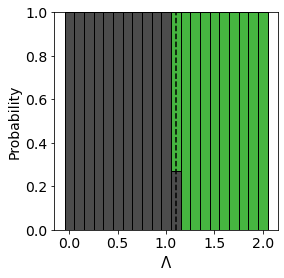}
        \end{tabular} \\
    \end{tabular}
%    \caption{Quantum phase prediction of the system via the CNN. The graphs show the probability that the system is on a given phase for each point, predicted only from the time evolution of the $C_z(1,2)$ correlation function; both the exact solution and the one obtained from the Trotter expansion with $n_T=6$ are presented for the initial state $|\downarrow_1 \ \downarrow_2 \ \downarrow_3 \ \downarrow_4 \ \uparrow_5 \ \uparrow_6 \ \uparrow_7 \ \uparrow_8 \ \rangle$ with $\varepsilon = 1$. QPTs are showed moving through the following lines of the phase space: \textbf{A)} $\Sigma = 0.5$ and $\Lambda = 0.5$ moving from $\chi = 0$ (Symmetric) to $\chi = 2$ (HF). \textbf{B)} $\chi = 0.5$ and $\Lambda = 0.5$ moving from $\Sigma = 0$ (Symmetric) to $\Sigma = 2$ (BCS). \textbf{C)} $\chi = 0.5$ and $\Sigma = 0.5$ moving from $\Lambda = 0$ (Symmetric) to $\Lambda = 2$ (Combined HF-BCS). \textbf{D)} $\chi = 1.5$ and $\Lambda = 0.5$ moving from $\Sigma = 0$ (HF) to $\Sigma = 2$ (BCS). \textbf{E)} $\chi = 1.5$ and $\Sigma = 0.5$ moving from $\Lambda = 0$ (HF) to $\Lambda = 2$ (Combined HF-BCS). \textbf{F)} $\chi = 0.5$ and $\Sigma = 1.5$ moving from $\Lambda = 0$ (BCS) to $\Lambda = 2$ (Combined HF-BCS). The dashed black line in each graph denotes the critical point between phases for each case.}
\caption{Same caption as Fig~\ref{fig:MLPpred} but using the CNN.}
    \label{fig:Convpred}
\end{figure*}

\subsection{Convolutional Neural Network}
\label{sec-CNN}
Convolutional Neural Networks (CNN) are a type of deep learning algorithms that are well suited for image recognition. Their efficacy for extracting features and learning to classify images has been proven in multiple instances \cite{CNNPedestrian,CNN701181,CNNTraffic}. In our case, instead of feeding 2D arrays of pixel values, 1D arrays containing the time dynamics of the correlation function will be used. When working with 2D CNN, the data have actually the shape $(x, y, z)$, being $z$ the value contained in the pixel, while $x$ and $y$ are its position. 1D CNN work effectively in the same way, just with data of shape $(x, y)$, being $y$ the value of the variable of interest at time $x$. The general structure of a CNN includes more than one convolution layer, so that one can extract higher level features. They also include additional layers, e.g., the pooling and the fully connected layers, each with a different purpose. For the present model the used layers are,
\begin{itemize}
\item \textbf{Convolution layer}: the layer responsible of performing the convolution operation.
  % Usually it is comprised of two parts: a kernel or filter and an activation function, though in this case the activation function is applied in a following layer separately to allow for higher customization. The convolution is performed by applying several filters to small sections of the data, which are comprised of 3 data points for this model. The filters are applied to different sections of the data until all of it is covered, activating the neurons of the CNN based on the applied filter and the activation function. The goal of the training is that the CNN will learn to optimize these filters to detect the features that best classify the data.
    
\item \textbf{Activation layer}: the layer that applies the activation function together with the filter of the convolution layer.
  %There are many activation functions that can successfully be used with CNNs. Variations of the Rectified Linear Unit (ReLU) are the most widely used for their various advantages over other activation functions, like the Leaky ReLU employed here. For an in-depth discussion on activation functions, see Ref. \cite{dubey2021Activation}.
    
\item \textbf{Pooling layer}: the pooling layer performs a dimension reduction of the data, collapsing data by connecting clusters of neurons to a single neuron each.
  % , serving as a way to both save computational resources and to extract only the most relevant features from the data. Two types of pooling can be performed, average pooling, which takes the mean of the data that is being collapsed, or max pooling, which takes the maximum value of the data that is being collapsed. Max pooling is most commonly used, since for image recognition it serves as a noise cancelling method, but here it would erase some relevant features. For that reason, average pooling was chosen.
    
\item \textbf{Dropout layer}: this optional layer temporarily deactivates, or \textit{drops out}, randomly selected training parameters from the previous layer that has trainable parameters. Its goal is to avoid the ``overfitting''.
  % For fully connected layers, this is a neuron, for convolutional layers, a single weight of a single filter. When training CNNs, neurons adjust their weights based on the features extracted from the convolution layer and specific neurons tend to specialize on specific features. This specialization can lead to the model being too good at remembering some of the data provided for training, leaving it prone to overfitting. Randomly deactivating neurons forces other neurons to ``step in" to handle the representation required for predictions for the missing neurons, which leads to a better generalization and a more reliable model. In the case of the convolution layers, a special kind of dropout layer, known as spatial dropout, was used. Spatial dropout discards whole filters, since deactivating random weights from a single filter tends to slow down the training process without preventing it from overfitting the data.
    
\item \textbf{Fully Connected layer}: Also know as \textit{dense} layers, they connect every neuron of the input to every neuron of the output.
  % Functionally, they are the same as an MLP. Adding more dense layers or more neurons to each of this layers makes the network more complex and generally more able to properly classify the input values as long as you have enough data. Nevertheless, adding too many will slow the training process without a significant increase in accuracy and eventually, a decrease in accuracy due to overfitting. 
    
\item \textbf{Softmax layer}: This layer is a fully connected or dense layer that applies a specific kind of activation function, called a \textit{softmax} function, which is a normalized exponential function.
  % This is the last step of the CNN, connecting the neurons activated in the last layer to the output neurons, which each represents one of the labels for classification (five in this case, for the five quantum phases of the system). Internally, the network uses cross-entropy loss as a scoring function to determine how good a classification is. This dense layer applies a normalized exponential function, the so-called softmax function, effectively transforming this scoring into a list of normalized probabilities for each of the classification labels. In other words, given the correlation function $C_z(1,2)$, the final output of the CNN shows the probability of the system being in each of the five quantum phases, with the sum of the probabilities for every quantum phase adding up to one, much like the output of the MLP.
\end{itemize}
\begin{table}
\begin{center}
\begin{tabular}{ | c | c | c | }
\hline
\textbf{Layer}  &  \textbf{Output shape}  &  \textbf{\# of parameters}  \\ 
\hline\hline
Convolution 1D (1) & 100 $\times$ 32 & 128 \\  
\hline
Leaky ReLU (1) & 100 $\times$ 32 & 0 \\
\hline
Avg. Pooling 1D (1) & 34 $\times$ 32 & 0 \\
\hline
Spatial Dropout (1) & 34 $\times$ 32 & 0 \\
\hline
Convolution 1D (2) & 34 $\times$ 64 & 6208 \\  
\hline
Leaky ReLU (2) & 34 $\times$ 64 & 0 \\
\hline
Avg. Pooling 1D (2) & 12 $\times$ 64 & 0 \\
\hline
Spatial Dropout (2) & 12 $\times$ 64 & 0 \\
\hline
Convolution 1D (3) & 12 $\times$ 128 & 24704 \\  
\hline
Leaky ReLU (3) & 12 $\times$ 128 & 0 \\
\hline
Avg. Pooling 1D (3) & 4 $\times$ 128 & 0 \\
\hline
Spatial Dropout (3) & 4 $\times$ 128 & 0 \\
\hline
Convolution 1D (4) & 4 $\times$ 256 & 98560 \\  
\hline
Leaky ReLU (4) & 4 $\times$ 256 & 0 \\
\hline
Avg. Pooling 1D (4) & 2 $\times$ 256 & 0 \\
\hline
Spatial Dropout (4) & 2 $\times$ 256 & 0 \\
\hline
Flatten &      512     &     0         \\
\hline
Dense (1)  &         512       &       262656  \\
\hline
Leaky ReLU (1) &  512    &          0    \\
\hline
Dropout (1)     &    512    &          0     \\
\hline
Dense (2)  &         512       &       262656  \\
\hline
Leaky ReLU (2) &  512    &          0    \\
\hline
Dropout (2)     &    512    &          0     \\
\hline
Dense (3)  &         512       &       262656  \\
\hline
Leaky ReLU (3) &  512    &          0    \\
\hline
Dropout (3)     &    512    &          0     \\
\hline
Dense (4)  &         512       &       262656  \\
\hline
Leaky ReLU (4) &  512    &          0    \\
\hline
Dropout (4)     &    512    &          0     \\
\hline
Dense (5)  &         512       &       262656  \\
\hline
Leaky ReLU (5) &  512    &          0    \\
\hline
Dropout (5)     &    512    &          0     \\
\hline
Softmax     &    5      &          2562  \\
\hline \hline
\textbf{Total parameters} & & \textbf{$1445445$} \\
\hline 
\end{tabular}
\end{center}
\caption{Number of parameters and shape of the output of the used convolution layers. }
%are applied with padding to match the size of the output to the size of the input for each of the applied filters; the shape of the output is then the size of the input times the number of applied filters (32, 64, 128 and 256 respectively for the first, second, third and fourth convolution layers). A leaky ReLU is applied as activation function for the convolution, with a leak of $\alpha = 0.1$. Max pooling is then applied, reducing the size of the input by a third (rounded up). The ``flatten'' layer reshapes the data into a single dimension array that is then fed to the fully connected dense layer. The final output is of size 5; five probabilities corresponding to the five quantum phases of the system.}
\label{tab:layers}
\end{table}

A fast and efficient way of implementing such a model is through the use of the \textit{Keras} library \cite{chollet2015keras}, a deep learning application programming interface built on top of the machine learning platform \textit{Tensorflow} \cite{tensorflow2015-whitepaper}. \textit{Keras} provides the ability to apply all the necessary layers sequentially, giving the model the final structure shown in table \ref{tab:layers}. Four cycles are applied, consisting on convolution, activation, pooling and dropout, allowing for a hierarchical decomposition of the input and the extraction of higher level features, as previously mentioned. Similarly, five additional fully connected (dense) layers are applied before the final output layer. The model includes a layer named ``Flatten''. This layer simply reshapes the output of the previous layer into a single dimension array so that it can be passed as the input to the dense layer.

Table \ref{tab:layers} shows, alongside the layers, the shape of the output and the number of trainable parameters of each layer. Note that only the convolution and dense layers have trainable parameters and the rest of the layers simply apply a chosen transformation to the data which is never changed throughout the training process. The convolution layers are applied with padding to match the first dimension of the output with the size of the corresponding input. The second dimension will be equal to the number of applied filters, $32$, $64$, $128$, and $256$ for the first, second, third and fourth convolution layers, respectively.
For convolution layers, the trainable parameters define each of the filters. Since the filters have a size $3$, each one will have three parameters or weights for each of the input channels, plus one bias weight. This number, multiplied by the number of applied filters, gives the total number of trainable parameters shown on the right column of Table \ref{tab:layers}. In the case of the dense layers, each connection has its own weight and all input neurons are connected to all output ones, so the total number of parameters will be the size of the input times the size of the output, plus one bias weight for each output neuron.

%%%%%%%%%%%%%%%%%%%%%%%%%%%%%%%%%%%%%%%%%%%%%%%%%%%%
%%%%%%%%%%%%%%%%%%%%%%%%%%%%%%%%%%%%%%%%%%%%%%%%%%%%

\begin{figure}
\begin{subfigure}
    \centering
    \includegraphics[scale=0.6]{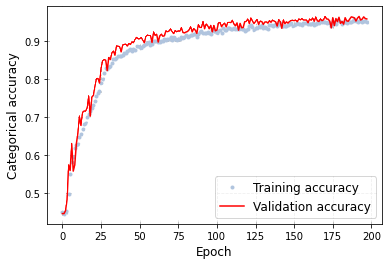}
    \caption{Training and test accuracy of the CNN as a function of the number of epochs used in the training process.} %The accuracy is a metric that measures the proportion of the data that the model correctly predicted. Validation data is not used in the backpropagation to compute the new weights of the neurons (that is, a training step), but as a benchmark to the generalization of the model, making sure it can predict data it has not use to train with the same accuracy as the training data. The model shows no signs of overfitting, as training and validation accuracy evolve at the same pace.}
    \label{fig:corracc}
\end{subfigure}
%\begin{subfigure}
%    \centering
%    \includegraphics[scale=0.6]{Figures/Convolution/CorrLoss.png}
%    \caption{Training and validation loss of the CNN. Loss is used as a scoring function for the model (the lower, the better the fit). %Discrepancies between the training and validation loss}
%    \label{fig:corrloss}
%\end{subfigure}
%\label{fig:accloss}
\end{figure}

Once the layer structure is defined, the model should be trained in a similar manner to the MLP case. The data are divided in three sets, namely, the test, the training and the validation set. The training one is used to get the trainable parameters, the validation one to check the consistency of the results and eventually to modify the network layer structure and hiperparameters (such as the number of neurons or filters), finally, the test dataset is used to report the prediction accuracy of the process. When working with Neural Networks, a common problem is the overfitting of the model, specially when the available data for training is limited. The model may ``memorize'' the training data and achieve good results for that input specifically, but becomes unable to generalize and predict correctly for data outside of the training set. In figure \ref{fig:corracc}, training and validation accuracy are shown across the epochs of the training process. In the training case, the accuracy in predicting the phase is computed using the training data, while in the validation accuracy a separate set of data is used, i.e., the validation set. Note that neither of these sets contain the test dataset. One can see that both training and validation accuracies steadily increase at a similar pace, which suggests that the model is not overfitted. In the case of overfitting, the training accuracy would increase faster than the validation accuracy \cite{oshea2015introduction} or the latter would stagnate at an earlier epoch, which is not the case. The total number of training epochs considered in this work was $200$, as further cycles do not significantly increase the accuracy, assuring that the model has converged.
In our case, we get an accuracy in the prediction of the phase, using the test dataset, of a $98.7\%$ (corresponding to the exact evolution). Note that this reported accuracy do not use data neither from the validation dataset nor from the validation one.
%because they are used to tune the hyperparameters of the CNN, and therefore, they are not valid to report the actual performance of the CNN. Reported accuracy was tested on a separate dataset (test data) which was not used at any moment during the training or optimization process. 

The increase in accuracy from $96.7\%$ using the MLP to $98.7\%$ using the CNN model is small, but significant. In particular, now the model has a better performance in the most complicated regions, i.e., near the QPTs. %Furthermore, the model performs decisively better for the desired task of exploring quantum phase transitions, which can be
This is clearly seen in Fig.~\ref{fig:Convpred}, especially for the cases b), c) and d) where the QPT becomes much more clear.
The main reason is that not only the model is more precise, but also that inaccuracies are mainly limited to points with values of $\chi$ and/or $\Sigma$ equal to or near zero, with only a few cases occurring at the critical points. The reason for these inaccuracies is that the amplitude of the correlation function is very small when both $\chi$ and $\Sigma$ are small, and as such, it presents very little variation as one moves across $\Lambda$. Hence, this inaccuracy is mostly present between the symmetric and the combined HF-BCS phases. In both cases, the time evolution is quite similar but the model seems to slightly favor the combined HF-BCS phase as the amplitude is generally lower in this case.

It is worth to mention once again, that the results are robust to the error introduced from the Trotter approximation even in the case of a relatively small number of Trotter steps. As a matter of fact, the model was trained and tested using the Trotterized time dynamics of the correlation function  $C_z(1,2)$ computed for $n_T=6$ and $n_T=20$ with accuracies of $99.2\%$ and $98.9\%$, respectively, to be compared with $98.7\%$ of the exact evolution.

As a conclusion of this section, we can say that the accuracy in determining the phase is larger for the CNN than for the MLP method, which is expected, specially in regions nearby the QPTs. Moreover, a really appealing fact is that the accuracy of the procedure is even larger when using the Trotter approximation with a small value of $n_T$, which has clear practical advantages. A possible explanation is that the larger oscillations observed in the approximate evolution, compared with the exact one, somehow helps the CNN to better predict the most problematic situations (see Appendix \ref{sec-app-A}). The use of a relatively small number of Trotter steps reduce the number of needed quantum gates and, therefore, will reduce the experimental error. 
%%%NEW
Because the above reasons, the practical implementation of this algorithm in a real quantum computer seems to be feasible and potentially valid to determine the phase of the system. As a matter of fact, the impact of noise on the learning accuracy, assuming it is small, it is not expected to affect the phase prediction.

\subsection{Further exploration}
  
Other correlation functions, as well as different initial states were studied. There is a lot of combinations of different possible correlation functions and different initial states to evolve.  %, so the study was not extensive. 
On the one hand, it is obvious that equivalent correlation functions will give the same results, in this case $C_z(1,2)=C_z(1,3)=C_z(2,4)=C_z(3,4)=C_z(5,6)=C_z(5,8)=C_z(6,8)=C_z(7,8)$ and their conjugates for the initial state $|\downarrow_1 \ \downarrow_2 \ \downarrow_3 \ \downarrow_4 \ \uparrow_5 \ \uparrow_6 \ \uparrow_7 \ \uparrow_8 \ \rangle$. On the other hand, the rest of the studied cases give comparable, although less accurate results. The CNN trained and tested on $C_z(1,4)$ for the initial state $|\downarrow_1 \ \downarrow_2 \ \downarrow_3 \ \downarrow_4 \ \uparrow_5 \ \uparrow_6 \ \uparrow_7 \ \uparrow_8 \ \rangle$ obtained an accuracy of $94.2\%$, while using the correlation function $C_z(1,2)$ for a different initial state $|\uparrow_1 \ \uparrow_2 \ \uparrow_3 \ \downarrow_4 \ \uparrow_5 \ \downarrow_6 \ \downarrow_7 \ \downarrow_8 \ \rangle$ achieved only an accuracy of $86.5\%$. Initial states that are eigenstates of one or more of the terms of the Hamiltonian are unsuitable as the information of the time evolution is partially (or totally) lost. Similarly, some correlation functions may vanish, for instance, $\sigma_x$ or $\sigma_y$ when using initial states in the z basis. Even when computing the time dynamics of the correlation function $C_x(1,2)$ for an initial state $|\downarrow_1^x \ \downarrow_2^x \ \downarrow_3^x \ \downarrow_4^x \ \uparrow_5^x \ \uparrow_6^x \ \uparrow_7^x \ \uparrow_8^x \ \rangle$ (where  $|\downarrow_i^x\rangle$ and $|\uparrow_i^x\rangle$ stand for the eigenvectors of $\sigma^x_i$ with eigenvalues $-1$ and $1$, respectively) which does not vanish, the CNN was unable of providing valuable results, i.e., it fails in finding the phases of the system.

At the moment, it is not evident how to discern which correlation function would be optimal to train the model and to classify the quantum phases of the system, apart of the visual inspection of the correlation function dynamics at different points in the phase space. Those with the most obvious differences to the naked eye seem to be also the ones that the CNN is able to better classify.

\section{Summary and conclusions} 
\label{sectVI}
In this work, an experimental setup for the quantum simulation of the extended Agassi model with $N = 8$ sites was proposed. The feasibility of this implementation with a Jordan-Wigner spin-mapping approach using trapped ions was studied. The estimations suggest that this protocol should be achievable in the near future as the fidelity of quantum gates will increase and new technologies will be developed. 

The rich quantum phase diagram of the extended Agassi model has been studied, showing that the time dynamics of a quantum correlation function, calculated with a particular initial state, can give information about the quantum phase of the system even though the quantum phase transitions regions are diffuse due to the finite size of the system. Two supervised machine learning models were trained to extract this information and work as classifiers of the quantum phase of the system, namely, a simple Multi-Layer Perceptron (MLP) and a more complex Convolutional Neural Network (CNN). The MLP has been proved to be an appropriate tool to explore the QPTs of the system, with an accuracy in determining the phase of the system of a $96.7\%$. %This is already enough to predict most of the phases of the system correctly and, specially, to be able to identify the QPTs of the system significantly better than with the progressive change seen by just visually inspecting the correlation functions. 
The CNN achieved a significantly better result with an accuracy of a $98.4\%$ in determining the phase and its performance is much better than the one of the MLP case to correctly classify the quantum phase near the critical points, giving a cleaner and more precise view of the quantum phase diagram. As previously mentioned, for $j=2$, finite $N$ effects make the differences between phases less apparent, but by increasing $j$ (and therefore $N$), the previously blurred QPTs become better defined. For that reason, it is remarkable that the machine learning procedure is capable of classifying the quantum phase of the system for a small size of the system $N$, as well as for a low value of $n_T$ Trotter steps. From an experimental point of view, keeping both quantities low makes the quantum simulation of the system much feasible with current technology.  %Furthermore, it was shown that, for higher number of sites, the QPTs of the Agassi model are much better defined, which once again proves that applying simple deep learning methods one can study the quantum phase of the system. Furthermore, one needs only to know the initial state and the time dynamics of quantities that can be easily measured experimentally, even when introducing experimental error from a protocol that relies on Trotterized dynamics.

%%%NEW
One must note that supervised machine learning algorithms require previous knowledge of the target system. In this case, the quantum phase diagram of the extended Agassi model is known thanks to mean field theory, and therefore, one can compute the necessary training data for the employed algorithms through an exact diagonalization in a classical computer. Naturally, cases where previous knowledge of the quantum phases is not available would be of special interest, for which different methods would be necessary. Despite limited by this fact, the presented approach serves as a proof of concept for future development, showing that machine learning algorithms can be powerful tools to extract quantum phase information from the time evolution of certain observables. It is then reasonable to think that, inspired by this idea, one could be able to obtain partial information of the phase diagram of an unexplored Hamiltonian using training data, in a more general manner, from known cases. Another option is to make use of unsupervised machine learning algorithms to deduce the quantum phases of a generic Hamiltonian without any prior knowledge of such phases, though this supposes a new set of challenges. Namely, unsupervised classification algorithms rely somehow on grouping the data accordingly on how similar they are, but cases near critical points are expected to present a similar behavior. Therefore, it is unlikely that popular unsupervised algorithms would be able to give a precise prediction of the position of the QPTs, but could still be useful to get a general qualitative idea of the quantum phase diagram of a model.

The use of the time evolution of a correlation function to determine the phase of the Agassi model was already introduced in Ref.~\cite{perezfernandez2021quantum}, but in the present work, its use has been improved through the implementation of machine learning models. It is worthy to mention that the phase of a system is a property of its ground state, however the time evolution of a given observable depends on the whole Hilbert space and on its energy spectrum. The quantum phase of the system strongly determine its spectrum and, therefore, will control the time evolution of any operator. An extreme case is the presence of excited state quantum phase transitions (ESQPTs) \cite{Capr08} in the spectrum which can be understood as a prolongation of the ground-state phase transition to the excited states and suppose a pileup of states at a particular energy, corresponding to a singularity in the density of states in the thermodynamic limit. The presence of a ESQPT in the spectrum has deep consequences on the time evolution of a non-eigenstate, as can be readily observed in works on the evolution of the fidelity \cite{Rela08} or the out-of-time-order correlators (OTOC) \cite{Wang19}.

\section*{Acknowledgements}

 This work was partially supported by the Consejer\'{\i}a de Econom\'{\i}a, Conocimiento, Empresas y Universidad de la Junta de Andaluc\'{\i}a (Spain) under Groups FQM-160, FQM-177, and FQM-370, and under projects P20-00617, P20-00764,  P20-01247, UHU-1262561, and US-1380840; by grants PGC2018-095113-B-I00, PID2019-104002GB-C21, PID2019-104002GB-C22, and PID2020-114687GB-I00 funded by MCIN/AEI/10.13039/50110001103 and ``ERDF A way of making Europe'' and by ERDF, ref.\ SOMM17/6105/UGR. Resources supporting this work were provided by the CEAFMC and Universidad de Huelva High Performance Computer (HPC@UHU) funded by ERDF/MI\-NE\-CO project UNHU-15CE-2848.

\bibliographystyle{IEEEtran}
\bibliography{references} % see references.bib for bibliography management
 
%\vspace{9pt} \hrule \vspace{3pt}

\appendix
\section{Time evolution of $C_z(1,2)$ for selected cases}
\label{sec-app-A}
\begin{figure}
\hspace{30pt} a) \hspace{110pt} b)
  \centering
  \includegraphics[width=0.5\textwidth]{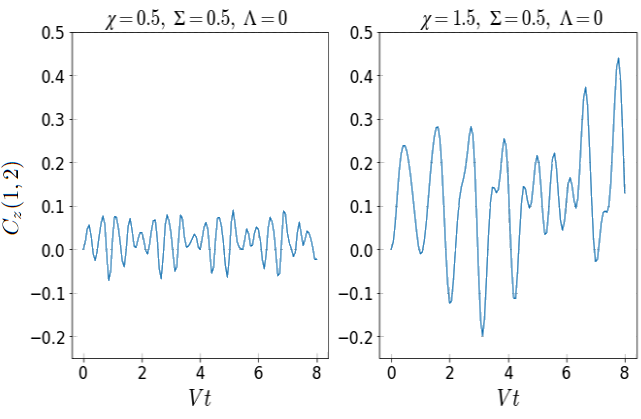}
  \caption{Time evolution of the correlation function $C_z(1,2)$ for $\chi = 0.5$, a), and $\chi=1.5$, b),with $\Sigma = 0.5$ and $\Lambda=0$ 
  %$\varepsilon=1$ 
  and an initial state $|\downarrow_1 \ \downarrow_2 \ \downarrow_3 \ \downarrow_4 \ \uparrow_5 \ \uparrow_6 \ \uparrow_7 \ \uparrow_8 \ \rangle$. %The system undergoes a phase transition from the Symmetric phase ($\chi < 1$, on the left) to the HF deformed phase ($\chi > 1$, on the right). After the QPT, the oscillations increase in amplitude and mean value, while decreasing in frequency.
  }
    \label{fig:C12HF}
\end{figure}

\begin{figure}
\hspace{30pt} a) \hspace{110pt} b)
    \centering
    \includegraphics[width=0.5\textwidth]{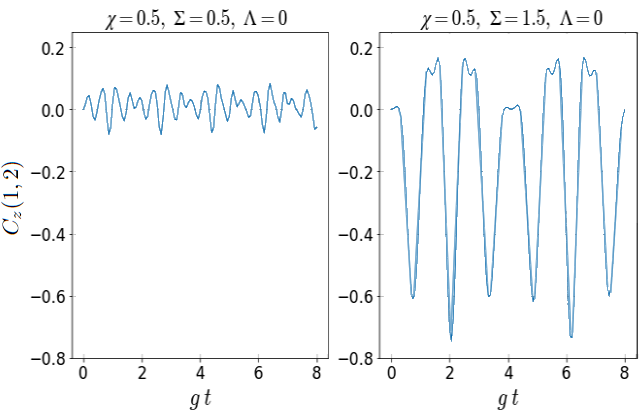}
    \caption{Same caption as Fig.~\ref{fig:C12HF} but with $\chi = 0.5$, $\Lambda=0$ and a) $\Sigma = 0.5$ and b) $\Sigma=1.5$. 
    %,  and initial state $|\downarrow_1 \ \downarrow_2 \ \downarrow_3 \ \downarrow_4 \ \uparrow_5 \ \uparrow_6 \ \uparrow_7 \ \uparrow_8 \ \rangle$. 
    %The system undergoes a phase transition from the symmetric phase ($\Sigma < 1$, on the left) to the BCS deformed one ($\Sigma > 1$, on the right). After the QPT, the oscillations increase in amplitude, but unlike the case for the HF deformed phase, the correlation function oscillates around negative values
    }
    \label{fig:C12BCS}
\end{figure}

\begin{figure}
    \hspace{30pt} a) \hspace{110pt} b)
    \centering
    \includegraphics[width=0.5\textwidth]{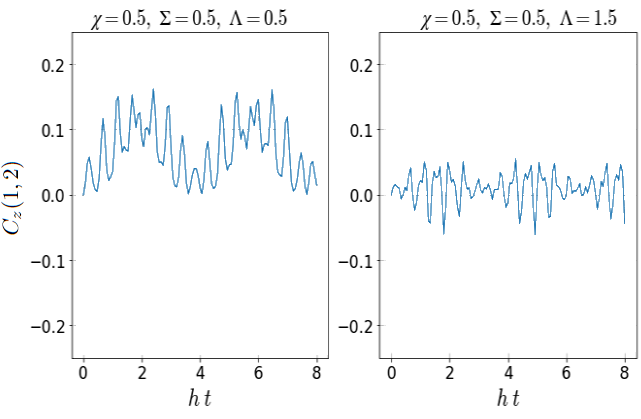}
    \caption{Same caption as Fig.~\ref{fig:C12HF} but with $\Lambda = 0.5$, a) and $\Lambda=1.5$, b), $\chi = 0.5$ and $\Sigma=0.5$.
    %, and initial state $|\downarrow_1 \ \downarrow_2 \ \downarrow_3 \ \downarrow_4 \ \uparrow_5 \ \uparrow_6 \ \uparrow_7 \ \uparrow_8 \ \rangle$. The system undergoes a phase transition from the symmetric phase ($\Lambda < 1$, on the left) to the BCS deformed one ($\Lambda > 1$, on the right). After the QPT, the amplitude of the oscillations decreases, while frequency remains approximately the same. The way the amplitude is modulated also changes.
    }
    \label{fig:C12Comb}
\end{figure}

%\begin{center}
%\textbf{APPENDIX \hypertarget{appendixB}{B}:}
%\end{center}
Time evolution of the correlation function $C_z(1,2)$ (\ref{sigma12}) for selected points of the phase diagram of the extended Agassi model. See Figs.~\ref{fig:C12HF}-\ref{fig:C12Comb}. It can be seen that the time dynamics of this correlation function is significantly different for each quantum phase of the system, even though the differences are difficult to be described quantitatively.%, which supports the idea of using machine learning models to capture these differences and to classify the quantum phase based on them.


\begin{figure}
    \centering
    \includegraphics[width=0.45\textwidth]{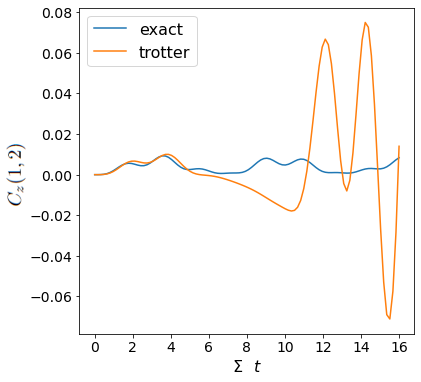}
    \caption{Time evolution of the correlation function $C_z(1,2)$ with parameters $\chi=0.0$, $\Sigma=0.1$ and $\Lambda=0.6$ using the exact evolution (in blue) and the Trotter expansion (in orange). 
    %, and initial state $|\downarrow_1 \ \downarrow_2 \ \downarrow_3 \ \downarrow_4 \ \uparrow_5 \ \uparrow_6 \ \uparrow_7 \ \uparrow_8 \ \rangle$. The system undergoes a phase transition from the symmetric phase ($\Lambda < 1$, on the left) to the BCS deformed one ($\Lambda > 1$, on the right). After the QPT, the amplitude of the oscillations decreases, while frequency remains approximately the same. The way the amplitude is modulated also changes.
    }
    \label{fig:TrotterVsExact}
\end{figure}

In Fig.~\ref{fig:TrotterVsExact}, it can be seen how the Trotter expansion differs from the exact evolution. The larger oscillations at higher times are a probable explanation for the enhanced performance of the CNN when trained and tested on the Trotterized data. Note that the error introduced from using the Trotter expansion is not arbitrary noise, it explicitly depends on the various terms of the Hamiltonian, so it still carries the information about the phase of the system. This error likely makes the features of the input data easier to recognize by the CNN than the much smoother exact evolution.

\end{document}